# A Fast Improved Fat Tree Encoder for Wave Union TDC in an FPGA[*]

SHEN Qi(沈奇)[1,2]　ZHAO Lei(赵雷)[1,2; 1)]　LIU Shu-Bin(刘树彬)[1,2]　LIAO Sheng-Kai(廖胜凯)[2,3]　QI Bin-Xiang(祁宾祥)[1,2]　HU Xue-Ye(胡雪野)[1,2]　PENG Cheng-Zhi(彭承志)[2,3]　AN Qi(安琪)[1,2]

[1] State Key Laboratory of Particle Detection and Electronics, University of Science and Technology of China, Hefei 230026, China

[2] Department of Modern Physics, University of Science and Technology of China, Hefei 230026, China

[3] Hefei National Laboratory for Physical Sciences at Microscale, Hefei 230026, China

**Abstract:** Up to the present, the wave union method can achieve the best timing performance in FPGA based TDC designs. However, it should be guaranteed in such a structure that the non-thermometer code to binary code (NTH2B) encoding process should be finished within just one system clock cycle. So the implementation of the NTH2B encoder is quite challenging considering the high speed requirement. Besides, the high resolution wave union TDC also demands the encoder to convert an ultra-wide input code to a binary code. We present a fast improved fat tree encoder (IFTE) to fulfill such requirements, in which bubble error suppression is also integrated. With this encoder scheme, a wave union TDC with 7.7 ps RMS and 3.8 ps effective bin size was implemented in an FPGA from Xilinx Virtex 5 family. An encoding time of 8.33 ns was achieved for a 276-bit non-thermometer code to a 9-bit binary code conversion. We conducted a series of tests on the oscillating period of the wave union launcher, as well as the overall performance of the TDC; test results indicate that the IFTE works well. In fact, in the implementation of this encoder, no manual routing or special constrains were required; therefore, this IFTE structure could also be further applied in other delay chain based FPGA TDCs.

[*] Supported in part by the Knowledge Innovation Program of the Chinese Academy of Sciences (KJCX2-YW-N27) and in part by the National Natural Science Foundation of China (11222552), and in part by the Fundamental Research Funds for the Central Universities (WK2030040015).

1)E-mail: zlei@ustc.edu.cn

# 1 Introduction

High resolution time measurement is very important in high energy physics (HEP) experiments, especially in time-of-flight (TOF) detector systems, and time-to-digital converters (TDC) are the essential parts in it. For instance, in the readout electronics of the a large ion collider experiment (ALICE) TOF detector [1], the kernel is an ASIC TDC chip named HPTDC which was developed specifically for HEP applications [2]. The HPTDC was also used in the electronics of the Beijing spectrometer (BESIII) TOF system [3], in which a 25 ps time resolution is achieved. There is a good review of the evolution of time-to-digital converters in Ref. [4]. Compared with ASIC TDCs, FPGA based TDCs have the benefits of shorter development time, and better flexibility. More and more TDCs have been implemented in FPGAs with the use of delay lines to perform time interpolation [5–9]. In FPGA TDCs, the final resolution is limited by the cell delays determined by the FPGA manufacturing process. The wave union TDC scheme first proposed by J. Wu in 2008 [10] can further improve both the effective bin size and timing precision (RMS) beyond the above limitation. Based on the wave union method, a typical timing performance of 9 ps RMS and 12 ps effective bin size was achieved in Xilinx Virtex 4 FPGA in our previous research [11].

However, the wave union method requires a high speed non-thermometer code to binary code (NTH2B) encoder to ensure that the TDC functions well. Once there is an input hit, the wave union launcher [11] with several 0-to-1 or 1-to-0 logic transitions will be triggered and fed into the TDC delay line. Then the states of the delay cells are latched on each rising edge of the following system clock cycles. Therefore, the encoder process must be finished within just one clock cycle. As the wave union launcher is essentially a ring oscillator [11], the raw bins are in the pattern as 111…1000…0111…1 or 000…0111…1000…0 which is actually a non–thermometer code. The delay time of the tapped delay cell of the FPGA is about tens of picoseconds. Considering a clock period of 10 ns, the number of delay cells, i.e. the length of the non-thermometer code, could be more than 200. Since the encoding process must be finished within just one clock cycle, it is quite challenging to implement the encoder to convert such an ultra-wide non-thermometer code to a binary code in such a short time period. The NTH2B encoder also needs to filter out the bubble errors in the non-thermometer codes, which are caused by different sources, e.g. meta-stability, uneven propagation delays, etc.

In our wave union TDC design, the period of the wave union launcher is designed to be longer than the system clock period so that there is no more than one oscillating rising edge in every clock cycle. This results in that the non-thermometer code is quite similar to the thermometer code, which can simplify the NTH2B encoding process. The methods used in thermometer code to binary code (TH2B) encoders can be references for the design of this NTH2B encoder. There has been much work in implementation of high speed and low power TH2B encoders in ASICs for flash analog to digital converters [12–17]. A common approach is to use a gray or binary ROM based encoder [14]. Although the ROM-based encoder is simple and straight forward to design, it is slow and cannot suppress bubble errors. The Wallace tree based encoder was first used by C. Wallace in 1964 [18], which counts the number of bit "1" in the thermometer code. The advantage of this approach is bubble suppression, however, accompanied by the disadvantage of a large delay and power consumption. A MUX based encoder was proposed in [19] which consists entirely of multiplexers; this encoder results in less complexity and a shorter critical path than Wallace tree encoder. However, it results in large fan-out in the critical path. In 2002, D. Lee, et al. proposed the fat tree encoder in ASICs [15]. This approach has the benefits of a high encoding speed and low power consumption. However, there have been few reports about TH2B encoder or NTH2B encoder implemented in FPGAs.

Inspired by the fat tree encoder method, we developed a fast improved fat tree encoder (IFTE) with special consideration for a wave union TDC implemented in Xilinx Virtex 5 XC5VLX110-1FFG676 FPGA device. The pipeline technique was introduced in our IFTE to overcome the unpredictable interconnection delays in the FPGA. With the special structure of the IFTE, the timing integrity can be easily satisfied with no manual placement and routing in the implementation, so the IFTE is device independent, which can be easily transplanted from one type of FPGA to another. Therefore, this IFTE structure could also be further applied in other delay chain based FPGA TDCs.

The outline of this paper is as follows. Section 2 describes the design and implementation of the proposed IFTE. Test results are presented in section 3 and conclusion is drawn in section 4.

## 2 Methodology and implementation

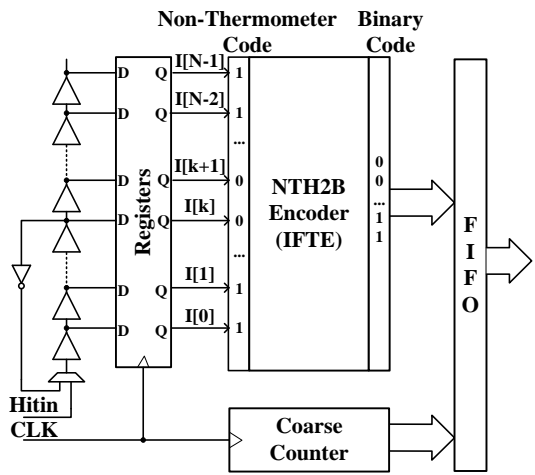

Fig. 1. Diagram of entire wave union TDC.

The block diagram of an entire wave union TDC is depicted in Fig. 1. The wave union launcher is embedded inside the tapped delay line. The IFTE imports the non-thermometer codes generated by the registers following the delay line, and converts them to binary codes. The binary codes are then written to the readout FIFO along with output of the coarse counter.

The encoding process is carried out in two stages in the IFTE, as shown in Fig. 2. The first stage converts the non-thermometer code to a one-out-of-N code, which is further converted by the second stage to a binary code using an improved fat tree structure.

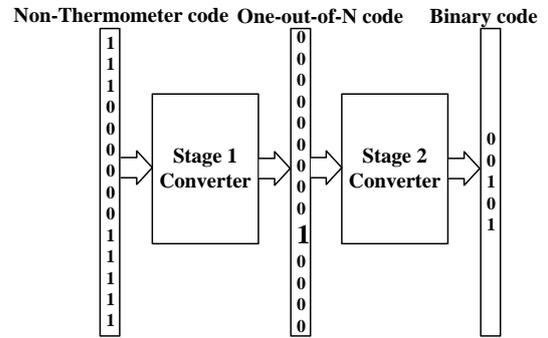

Fig. 2. The two-stage IFTE.

### 2.1 Non-thermometer code to one-out-of-N code converter (NT2ONC)

#### 2.1.1 Bit patterns of the raw bins

In the implementation of our FPGA TDC, the length of the tapped delay line ($T_{TDL}$) is longer than the system clock period ($T_{CLK}$) to allow for the tapped delay variation due to temperature and power supply voltage. As mentioned above, the oscillating period of the wave union launcher ($T_{OSC}$) is designed to exceed $T_{TDL}$ so that there is no more than one valid oscillating edge in each CLK cycle, in which the valid edge is defined as 1-0 transition from right to left in the raw bins I[(N-1):0] (for example, ..000111...).

The relationship among $T_{OSC}$, $T_{TDL}$, and $T_{CLK}$ is shown in Eq. (1).

$$T_{OSC} > T_{TDL} > T_{CLK}. \quad (1)$$

There are four bit patterns for the raw bins

I[(N-1):0] as shown in Fig. 3: (a) Pattern1 with one transition: I[(N-1):0]: 00…0011…11; (b) Pattern2 with one transition: I[(N-1):0]: 0…01…10…0; (c) Pattern3 with no transition: I[(N-1):0]: 11…1100…00; (d) Pattern4 with one transition: I[(N-1):0]: 1…10…01…1.

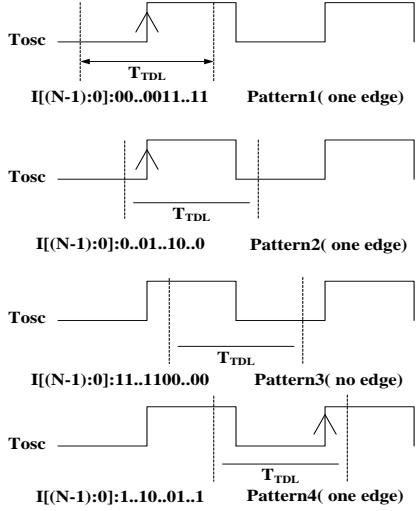

Fig. 3. Four bit patterns of raw bins.

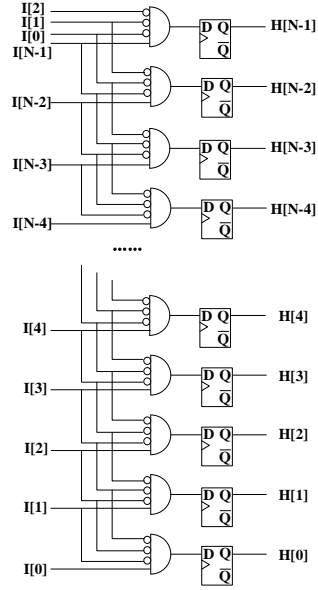

Fig. 4. Non-thermometer code to one-out-of-N code converter.

The function of the NT2ONC is to detect the 1-0 transition and record the position, and from this point of view, the non-thermometer code can partly be considered as a thermometer code. There is only at most one logical "1" appearing at the output of one-of-N code. The NT2ONC is implemented by using 4-input AND gates with 3 input inverted followed by D-flip-flops (DFFs), as shown in Fig. 4.

The one-out-of-N code ($H[(N-1):0]$) is calculated as Eq. (2). From Eq. (2) we can see that when Pattern3 with no transition occurs, i.e., $I[(N-1):0]=11..1100..00$, then the MSB of the one-out-of-N code $H[N-1]$ is "1" with all other bits equal to "0". This one-out-of-N code will be later converted to a binary coded maximum value of N-1, which corresponds to the length of the entire tapped delay line $T_{TDL}$. As $T_{TDL}$ exceeds the fine time range $T_{CLK}$, this maximum binary code makes no sense but just functions as a flag of Pattern3.

$$H[i] = \begin{cases} I[i] \& (\sim I[i+1]) \& (\sim I[i+2]) \& (\sim I[i+3]), & 0 \leq i \leq N-4 \\ I[i] \& (\sim I[i+1]) \& (\sim I[i+2]) \& (\sim I[0]), & i = N-3 \\ I[i] \& (\sim I[i+1]) \& (\sim I[0]) \& (\sim I[1]), & i = N-2 \\ I[i] \& (\sim I[0]) \& (\sim I[1]) \& (\sim I[2]), & i = N-1 \end{cases} \quad (2)$$

2.1.2 Bubble error suppression

The bubble errors around the transition edges may occur due to uneven propagation delays or noises in the FPGA structure. The common method to detect the 1-0 transition in the thermometer code is using a 2-input AND gate with one input inverted [14]. But 2-input AND gates cannot suppress the bubble error. In our design, when a bubble error occurs in the non-thermometer code, there will be more than one "1" appears at the one-out-of-N code, which may lead to catastrophic errors in the binary code. The 4-bit input AND gate recognizes a 4-bit pattern "0001" (not a 2-bit pattern "01") as a valid transition, so the one bit bubble cases, such as 0000101111(original: 0000111111), and the two bits bubble cases, such as 0000100111(original: 0000111111), could be eliminated. In our circuit, bubble errors with more than two bits [20] rarely occur so that the 4-bit input AND gate structure used here for transition detecting can eliminate the bubble error problems.

## 2.2 One-out-of-N code to binary code converter (ON2BCC)

The one-out-of-N code from the NT2ONC is passed to the second stage of the IFTE, One-Out-of-N code to binary code Converter (ON2BCC), in which multiple trees of OR gates and pipeline registers are employed. As mentioned above, the fat tree encoder was first used in an ASIC [15] with no need of pipeline registers, because the signal logic delays of OR gates from the inputs to binary code outputs are uniform, which is guaranteed by manual layout. However, the interconnection route delays in FPGAs are unpredictable, so special care must be taken. The total signal delays in FPGAs (consisting of the logic delays and route delays) of some critical paths could be much longer than the others; therefore, we added pipeline registers to guarantee a good timing margin in high speed situation.

We proposed a new IFTE structure with pipeline registers inserted in the fat tree encoder logic to reduce critical path delays. Fig. 5 shows an example of this IFTE structure with a 4-bit binary outputs case. The IFTE structure includes a basic tree and several output bit trees. A 16-bit one-out-of-N code H[15:0] is presented to the left nodes of the basic tree and a 4-bit binary code B[3:0] is produced at the right nodes of the output bit trees.

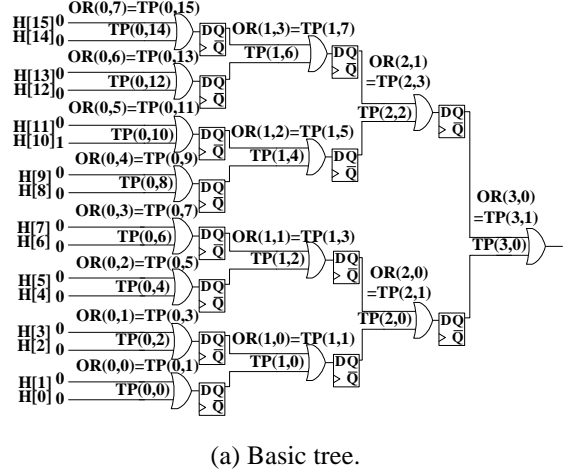

(a) Basic tree.

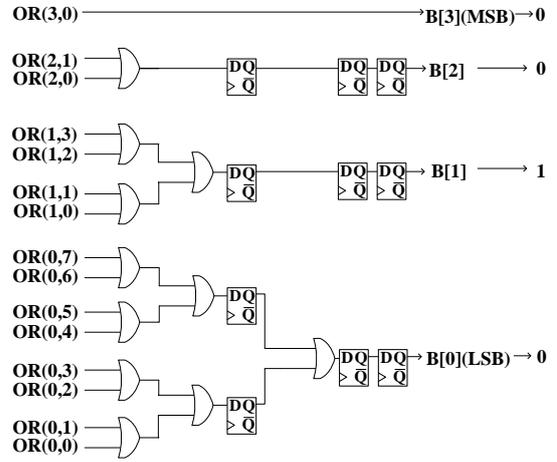

(b) Output bit trees.

Fig. 5. Example of the IFTE structure with a 4-bit binary outputs case.

The basic tree produces the intermediate signals TP(k,i) and OR(k,i) from the one-out-of-N code H[N-1:0] as in Eq. (3) and Eq. (4),

$$TP(k, i) = \begin{cases} H[i] & k = 0 \quad 0 \leq i \leq (N-1), \\ TP(k-1, 2i) + TP(k-1, 2i+1) & 1 \leq k \leq (n-1) \quad 0 \leq i \leq (\frac{N}{2^k} - 1), \end{cases} \quad (3)$$

$$OR(k, i) = TP(k, 2i + 1) \quad 0 \leq k \leq (n-1) \quad 0 \leq i \leq (\frac{N}{2^{k+1}} - 1), \quad (4)$$

where N and n represent the bit numbers of one-out-of-N code and the binary code, with a relation of $N = 2^n$. Then the output bit trees produce the binary code B[(n-1):0] from the OR(k,i) as in Eq. (5).

$$B[k] = \sum_{i=0}^{\frac{N}{2^{k+1}}-1} OR(k,i) \quad 0 \leq k \leq (n-1). \quad (5)$$

The pipeline register is inserted after every 2-input OR gate in the basic tree. In the output bit trees, the 2-input OR gates are replaced by 4-input OR gates as many as possible to reduce the amount of registers. And the 4-input OR gates are implemented with the use of the look-up-tables (LUTs) in the Virtex 5 FPGA.

A benefit of this IFTE structure is that all the binary output bits are processed in parallel which leads to a shorter encoding time. There exist n-1 pipeline stages from the N-bit one-out-of-N code to the MSB of the n-bit binary code in the basic tree. To uniform the latency of all binary output bits, additional registers are placed following the last stages of other output bit trees.

In our wave union TDC design, the system clock frequency is 120 MHz, i.e., the clock period is 8.33 ns, which corresponds to about 268 tapped delay cells. The raw non-thermometer code is designed to be 276 bits wide, so the length of the one-out-of-N code is also 276. We set N=512 and n=9 to satisfy the equation $N = 2^n$. The high bits of the one-out-of-N code H[511:276] are assigned constant values "0". In the implementation of the IFTE circuit, the Synthesize and Place &Route process are both automatically done by the software ISE with no manual placement and routing, in which the timing integrity is easily satisfied.

## 3 Test results

An evaluation board with readout via the PCI extensions for instrumentation (PXI) bus was designed to test the performance of the proposed IFTE scheme and the 2-channel wave union TDC implemented in the FPGA. The photograph of the board is shown in Fig. 6. The test results are presented as follows.

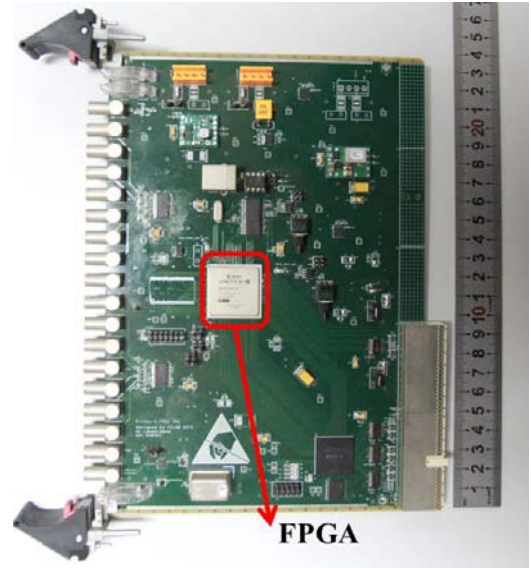

Fig. 6. Picture of the evaluation board.

## 3.1 Measurement of the oscillating period of the wave union launcher

The measurement of the oscillating period Tosc of the wave union launcher is used to estimate the performance of the encoder scheme of IFTE.

A periodic hit signal from a pulse generator, which is free running with respect to the TDC system clock, is fed to a TDC channel. Once there is a valid hit signal, the system clock samples the states of delay cells turn by turn for K cycles and the corresponding raw bits are recorded. As mentioned above, there exists no more than one wave union transition in each cycle, and K is set to be 16 in the design. The Tosc is calculated as Eq. (6) [11], in which $t_i$ and $t_{i+1}$ represent two adjacent fine times of the raw bits, i.e., the output binary codes of the IFTE. The Tosc can be directly obtained from the time interval of two adjacent oscillating transitions. Statistical analysis of Tosc is performed. With a large amount of test data, the Tosc will follow a normal distribution with a constant mean value. If the IFTE doesn't work well, error values of $t_i$ will be caused, which will then result in error values of Tosc. A statistical analysis of Tosc is performed for channel 1 with more than 100 thousands of data points. A typical result is shown in Fig. 7 (a), in which the mean value of Tosc is 9491 ps with a standard deviation of 31.3 ps. There is no error value of Tosc is observed, which indicates that the encoder scheme of this IFTE performs correctly and steadily.

$$T_{OSC}(i) = t_{i+1} - t_i + T_{CLK}, (1 \leq i \leq K - 1). \quad (6)$$

In order to verify the above method, Tosc is also measured with a LeCroy oscilloscope WavePro 715Zi (20GS/s with an analog bandwidth of 1.5GHz) by tying the output of the last tap of the delay chain to an FPGA output test pin. Fig. 7 (b) shows the graph of the Tosc, in which the mean value of Tosc is 9484 ps with a standard deviation of 21ps, which concords very well with the former test results. This further demonstrates the correctness of the IFTE encoder.

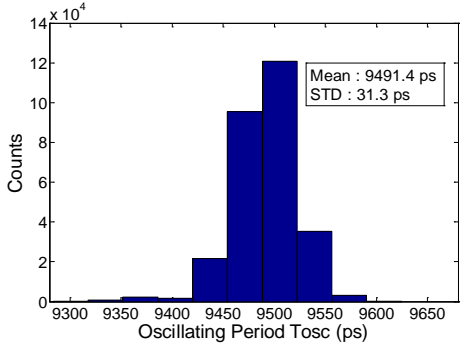

(a) Statistical spread of Tosc derived from the IFTE output codes.

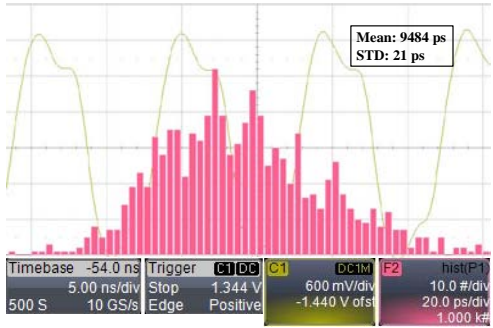

(b) Graph of Tosc measured with oscilloscope.

Fig. 7. Test Results of the oscillating period Tosc of the wave union launcher.

## 3.2 Performance of the wave union TDC

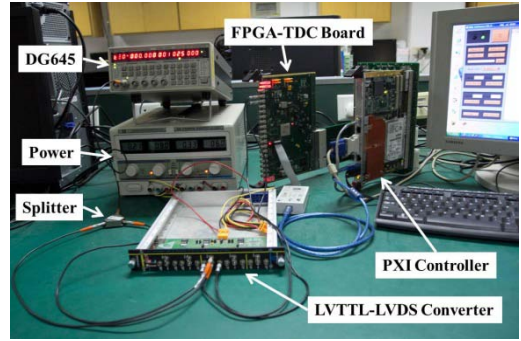

Fig. 8. The wave union TDC board under test.

We also conducted tests on the overall performance of the wave union TDC with the proposed fast IFTE integrated inside. As shown in Fig. 8, the cable delay test approach [21] is used to evaluate the overall precision. A pulse generator (DG645) is employed to produce a hit signal with a repetition frequency of 77.8 Hz, which is uncorrelated with the TDC clock. The hit signal is connected to two TDC channels after a splitter with a fixed cable delay between the channels. Then the time interval of the cable delay is measured by the two TDC channels. Fig. 9 shows the typical statistical spread of the cable delay measurement results with different amounts of wave union transitions. The standard deviation of the measured time interval ($RMS_M$) represents the system precision, which contains two TDC channels, hence a single TDC channel precision $RMS_{TDC}$ equals $RMS_M/\sqrt{2}$. The $RMS_{TDC}$ timing precision is reduced to 7.7 ps from 14.3 ps with the transition amounts increased to 8 from 1.

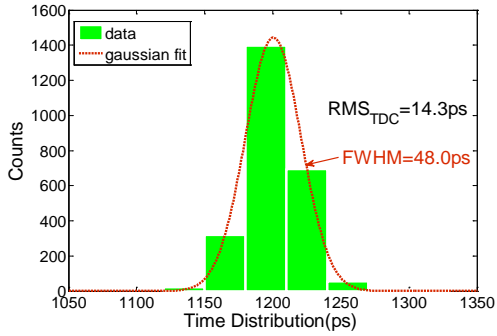

(a) RMS of plain TDC with 1 transition.

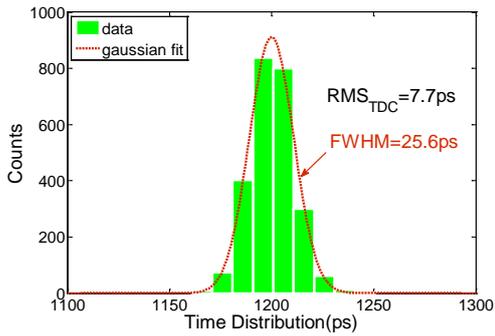

(b) RMS of wave union TDC with 8 transitions.

Fig. 9. Statistical spread of the cable delay test with different amounts of wave union transitions.

The effective bin size of the TDC can also be reduced by the wave union technique [11]. As seen in Fig. 10, the bins are effectively subdivided as the transition amounts increased. The binary codes are distributed to 268 distinct bins (Bin0 - Bin267) with 1 transition. As the system clock period $T_{CLK}$ is 8333.3 ps, the mean bin size can be calculated to be about 31 ps ($BIN_1 = T_{CLK}/268 = 31$ ps). For 4 transitions and 8 transitions, the effective bin size is reduced to 7.7 ps and 3.8 ps, respectively.

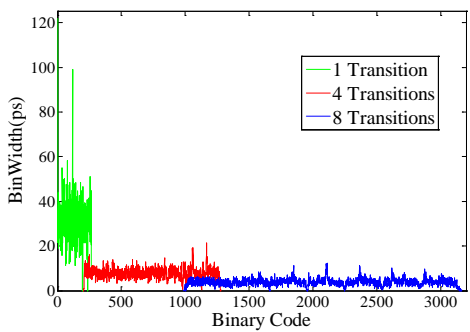

Fig. 10. Effective bin width plot of different times of wave union transitions.

## 4 Conclusion

We proposed a fast improved fat tree encoder, which is especially suitable for high speed, ultra wide non-thermometer code to binary code conversion. To verify the effectiveness of this method, a two-channel wave union TDC was designed, in which a 276-bit non-thermometer code to 9-bit binary code encoder was implemented with a clock cycle of 8.33 ns. Test results indicate that the encoder functions well and stably, and an overall RMS timing precision of 7.7 ps is achieved. In fact，with the high speed, bubble error suppression ability, and good transportability, this encoder could be also used in other delay chain-based FPGA TDCs.